\documentclass[3p,times,procedia]{elsarticle}
\usepackage{nupha_ecrc}
\usepackage{caption}
\usepackage{subcaption}

\volume{00}

\firstpage{1}

\journalname{Nuclear Physics A}

\runauth{}

\jid{nupha}

\jnltitlelogo{Nuclear Physics A}

\usepackage{amssymb}
\usepackage{lineno}

\usepackage[figuresright]{rotating}

\newcommand{\dndeta}       {\ensuremath{\mathrm{d}N_\mathrm{ch}/\mathrm{d}\eta}}

\newcommand{\dNdeta}       {\ensuremath{\langle\dndeta\rangle}}
\newcommand{\avNpart}      {\ensuremath{\langle N_\mathrm{part} \rangle}}
\newcommand{\dNdetape}     {\ensuremath{\frac{2}{\avNpart}\dNdeta}}

\begin{document}

\begin{frontmatter}

%% Title, authors and addresses

%% use the tnoteref command within \title for footnotes;
%% use the tnotetext command for the associated footnote;
%% use the fnref command within \author or \address for footnotes;
%% use the fntext command for the associated footnote;
%% use the corref command within \author for corresponding author footnotes;
%% use the cortext command for the associated footnote;
%% use the ead command for the email address,
%% and the form \ead[url] for the home page:
%%
%% \title{Title\tnoteref{label1}}
%% \tnotetext[label1]{}
%% \author{Name\corref{cor1}\fnref{label2}}
%% \ead{email address}
%% \ead[url]{home page}
%% \fntext[label2]{}
%% \cortext[cor1]{}
%% \address{Address\fnref{label3}}
%% \fntext[label3]{}

%% Instructions from Editor: Please use the following \dochead only in the preprint version (e-print arXiv etc.);
%% use empty \dochead{} when submitting to Nuclear Physics A!
\dochead{XXVIIth International Conference on Ultrarelativistic Nucleus-Nucleus Collisions\\ (Quark Matter 2018)}
%\dochead{}
%% Use \dochead if there is an article header, e.g. \dochead{Short communication}
%% \dochead can also be used to include a conference title, if directed by the editors
%% e.g. \dochead{17th International Conference on Dynamical Processes in Excited States of Solids}

\title{ALICE results on system-size dependence of charged-particle multiplicity density in p--Pb, Pb--Pb and Xe--Xe collisions}

%% use optional labels to link authors explicitly to addresses:
%% \author[label1,label2]{<author name>}
%% \address[label1]{<address>}
%% \address[label2]{<address>}

\author{Beomkyu Kim\\ (for the ALICE Collaboration)}

\address{INHA University, 100 Inha-ro, Michuhol-gu, Incheon 22212, KOREA}

\begin{abstract}
%% Text of abstract
Particle production at LHC energies involves the interplay of hard (perturbative)
and soft (non-perturbative) QCD processes. Global observables, such as the
charged-particle multiplicity, are related to the initial geometry and the energy
density produced in the collision. They are important to characterise
relativistic heavy-ion collisions and to constrain model calculations. The LHC
produced Xenon--Xenon collisions for the first time in October 2017. New
results on the primary charged-particle pseudorapidity density, and its
pseudorapidity and centrality dependence are presented for this lighter and
deformed nucleus, and compared to measurements obtained for lead ions. New
results will also be presented for p–-Pb collisions at the highest energy of 8.16
TeV, as part of an overview of all the measurements at LHC Run 1 and 2
energies. These studies allow us to investigate the evolution of particle
production with energy and system size and to compare models based on various
particle production mechanisms and different initial conditions.
\end{abstract}

\begin{keyword}
%% keywords here, in the form: keyword \sep keyword

%% MSC codes here, in the form: \MSC code \sep code
%% or \MSC[2008] code \sep code (2000 is the default)

ALICE \sep LHC \sep Xe--Xe \sep Pb--Pb \sep p--Pb \sep pp \sep Charged-particle multiplicity density \sep System-size dependence
\end{keyword}

\end{frontmatter}

%\linenumbers
%%
%% Start line numbering here if you want
%%
% \linenumbers

%% main text
\section{Introduction}
\label{}

For the last 8 years from 2010 to 2018, the ALICE Collaboration has provided
results of primary-charged-particle production in various collision energies and
systems. In October 2017, LHC collided xenon ions at
$\sqrt{s_\mathrm{NN}} = 5.44$ TeV.
Central collisions of heavy ions, like Pb (atomic number $\mathrm{A}=208$)
revealed Quark-Gluon Plasma (QGP) effects. Xe has fewer nucleons ($\mathrm{A}=129$)
than Pb and is a good medium-sized ion to check how the
system size of a colliding system relates to the creation of the hot and
dense medium. On the other hand, QGP-like effects
have been observed even in pp and p--A collisions, the so-called small systems~\cite{Abelev:2012ola,Aad:2012gla,Khachatryan:2015lva}.
In high multiplicity
pp and p--Pb collisions, the process of the Multi Parton Interactions (MPI)~\cite{Sjostrand:1987su} becomes more important
and is supposed to be related to QGP-like effect~\cite{Ortiz:2013yxa}. Particle production
at few GeV/$c$ is dominated by soft QCD and makes a big contribution to the charged-particle multiplicity density.
This can be approached by phenomenological modelling.

\section {Analysis method}
%The results presented here are a selection of measurements that  ALICE has collected
%from 2010 to 2018 in p--Pb, Pb--Pb and Xe--Xe collisions.
Primary charged-particle multiplicity density, $\mathrm{d}N_\mathrm{ch}/\mathrm{d}\eta$,
is measured by counting the number of tracklets (a short track segment)
using the SPD detector~\cite{1748-0221-4-03-P03023} in the central region ($-1.8<\eta<1.8$)
and estimating the effective energy density per charged particle using the FMD detector~\cite{Christensen:2007yc,Cortese:781854}
in the forward regions ($-3.5<\eta<-1.8$ and $1.8<\eta<5$). Data were collected with a
minimum bias trigger requiring a coincidence of signals in each side of V0
sub-detectors (V0A and V0C)~\cite{Abbas:2013taa, Cortese:781854}. The primary interaction vertex
of a collision is obtained by extending correlated hits in the two
silicon-pixel layers of the SPD to the beam pipe of the LHC.
%Multiplicity percentile in pp collisions
%is estimated by measuring the sum of amplitudes in the V0A and V0C detectors.
Centrality estimation is based on a Glauber
approach~\cite{Alver:2008aq,Loizides:2014vua} by fitting to the V0 amplitude
distribution~\cite{PhysRevC.88.044909} for the Pb-going side (V0A) in p--Pb collisions or for both A-going sides
(V0A and V0C) in Pb--Pb and Xe--Xe collisions.
The amplitude of the V0 detector
is fitted with a two-component model given by $N_\mathrm{sources} = f\times
N_\mathrm{part} + (1-f)\times N_\mathrm{coll}$ where $f$ fixes the relative
contributions of $N_\mathrm{part}$ (the number of participating nucleons taking
part in the collision) and $N_\mathrm{coll}$ (the number of binary collisions
among participating nucleons), and $N_\mathrm{sources}$ is the number of intermediate sources
for the particle production to be transformed to the charged-particle multiplicity
by the negative binomial distribution (NBD).
The detailed systematic studies
in p--Pb, Pb--Pb and Xe--Xe collisions can be found
elsewhere~\cite{Acharya:2018hhy,Adam:2014qja,Aamodt:2010cz,Adam:2015ptt}.

\section{Results and discussion}

\begin{figure}[h]
  \begin{subfigure}[t]{0.47\textwidth}
      \includegraphics[width=\linewidth]{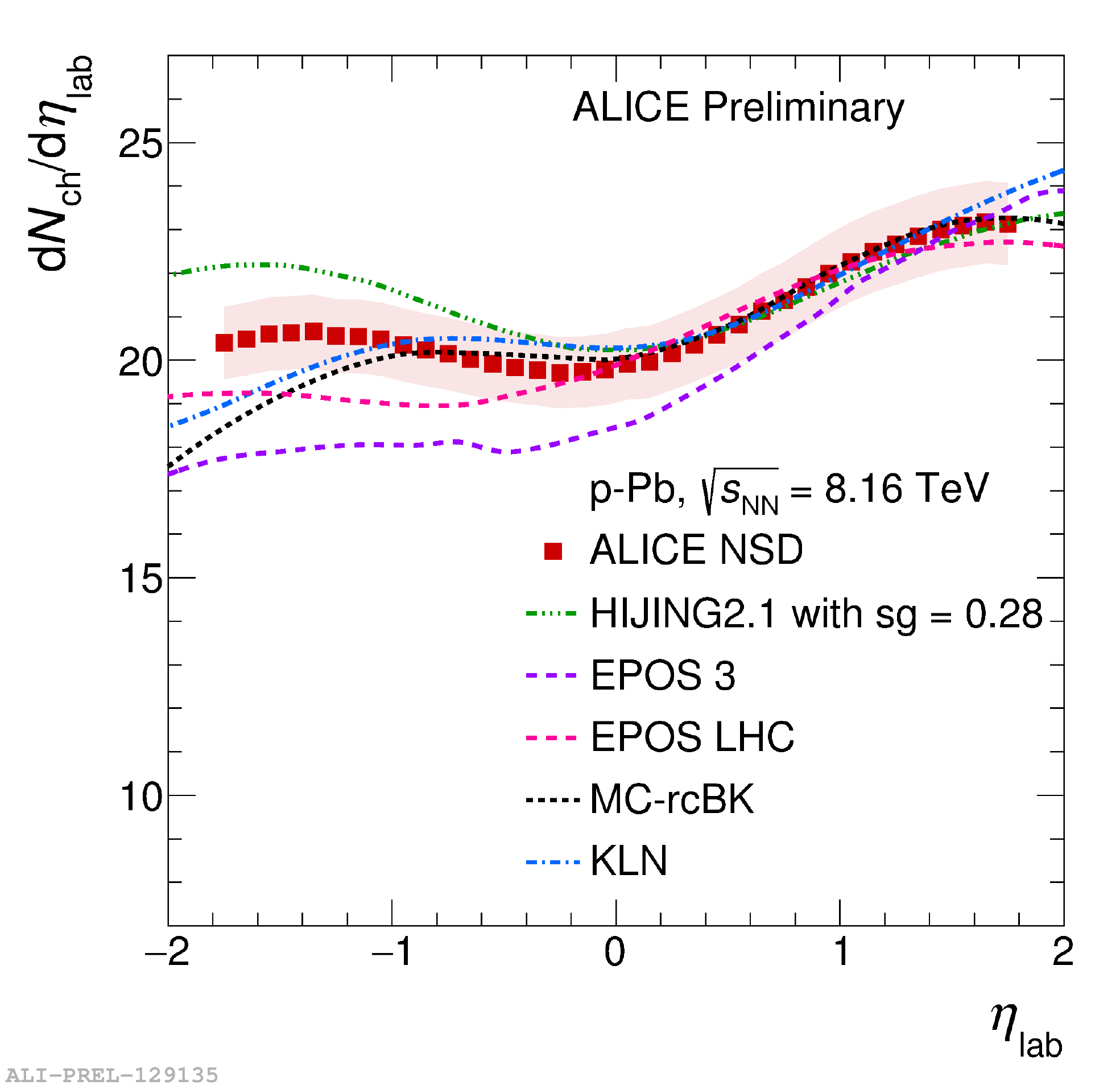}
      \caption{\dndeta\ in p--Pb collisions}
      \label{fig:plead}
    \end{subfigure}
  \begin{subfigure}[t]{0.47\textwidth}
      \includegraphics[width=\linewidth]{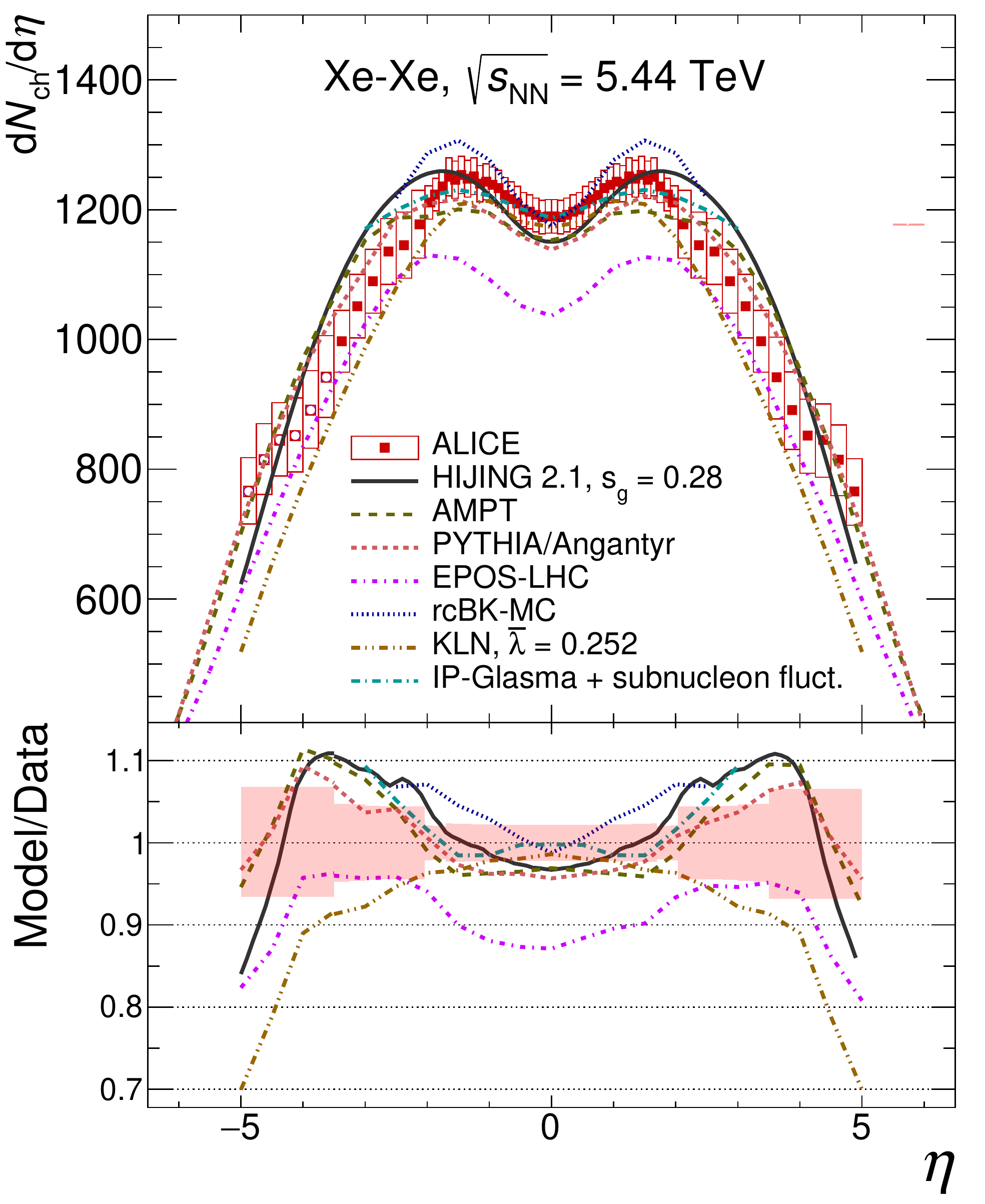}
      \caption{\dndeta\ for the top 5\% central Xe--Xe collisions~\cite{Acharya:2018hhy}}
      \label{fig:xexedndeta}
    \end{subfigure}
\end{figure}

Figure \ref{fig:plead} shows \dndeta\ in p--Pb collisions at
$\sqrt{s_\mathrm{NN}}=8.16$ TeV and the result is compared with theoretical
models. Phenomenological models like HIJING~\cite{Xu:2012au,PhysRevC.83.014915}, EPOS LHC~\cite{PhysRevC.92.034906}
and EPOS 3~\cite{Porteboeuf:2010um} reproduce the data
better for the Pb-going side than for the proton-going side. Saturation-based models like
rc-BK~\cite{Albacete:2012xq,Albacete:2010ad} and KLN~\cite{PhysRevC.85.044920}
describe the data better in $\eta_\mathrm{lab}>-1.3$ as expected.
However, all models describe the data within $\pm15\%$.
The \dndeta\ for the top $5\%$ central
Xe--Xe collisions is shown and compared to models in Fig. \ref{fig:xexedndeta}.
The pseudorapidity dependence of \dndeta\ is described by the models within $\pm10\%$ in $|\eta|<4$ except for EPOS LHC.

\begin{figure}[h]
  \begin{center}
    \begin{subfigure}[t]{0.47\textwidth}
      \includegraphics[width=\linewidth]{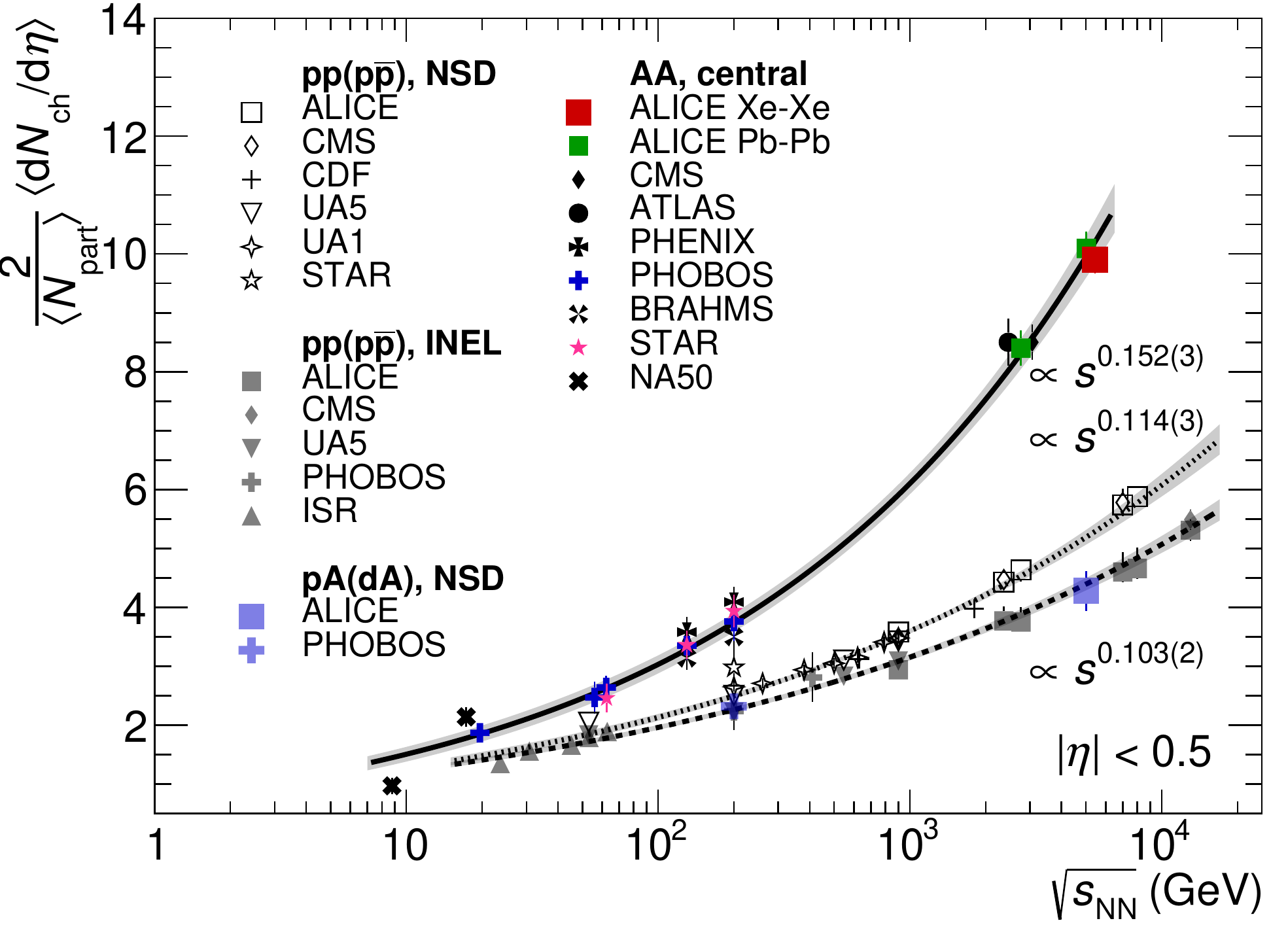}
      \caption{$\dNdetape$ vs $\sqrt{s}$, $\sqrt{s_\mathrm{NN}}$ ~\cite{Acharya:2018hhy}}
      \label{fig:roots}
    \end{subfigure}
    \begin{subfigure}[t]{0.47\textwidth}
      \includegraphics[width=\linewidth]{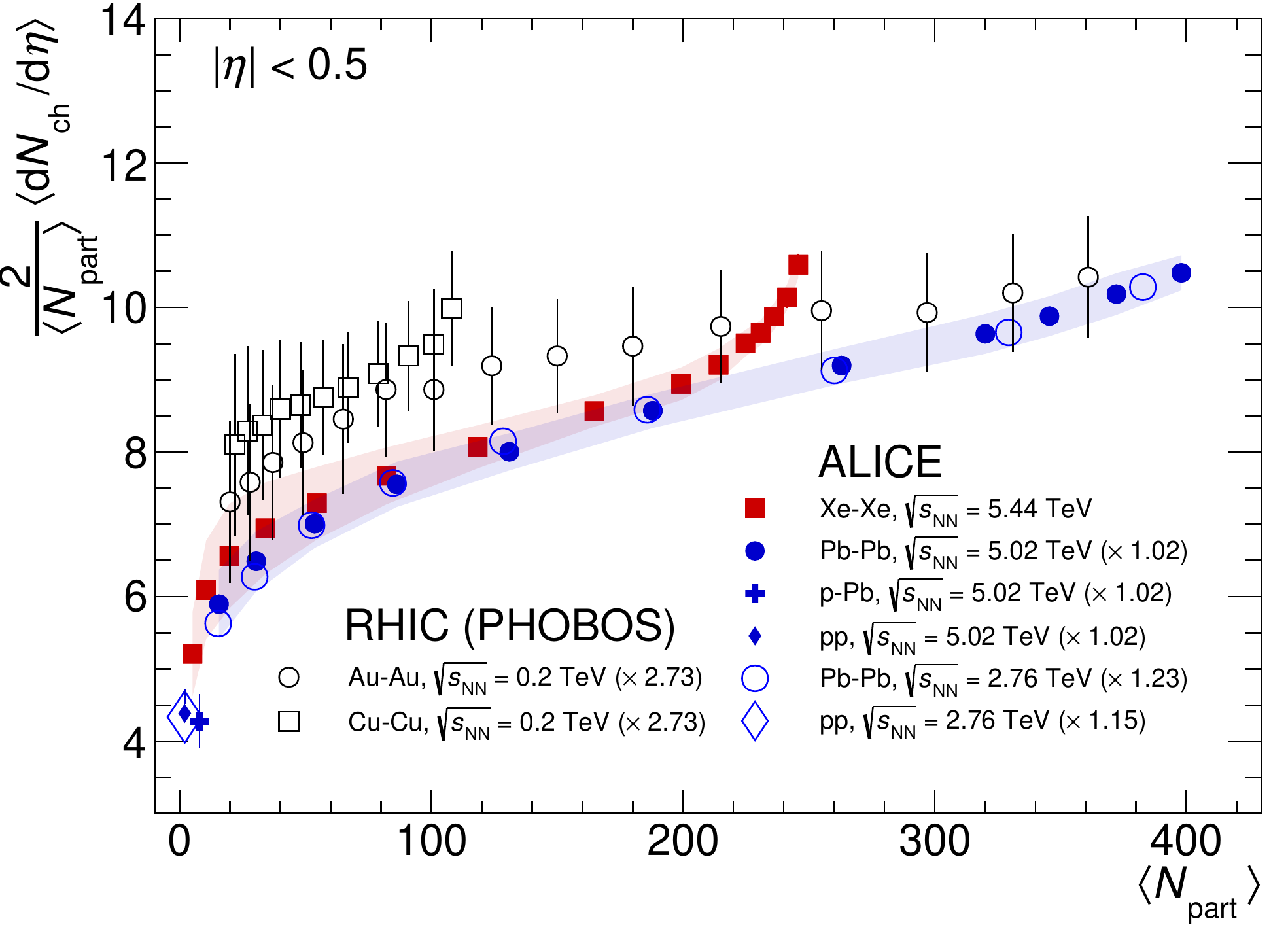}
      \caption{The \dNdetape\ vs \avNpart~\cite{Acharya:2018hhy}}
      \label{fig:npart}
    \end{subfigure}
    \begin{subfigure}[t]{0.47\textwidth}
      \centering
      \includegraphics[width=0.9\linewidth]{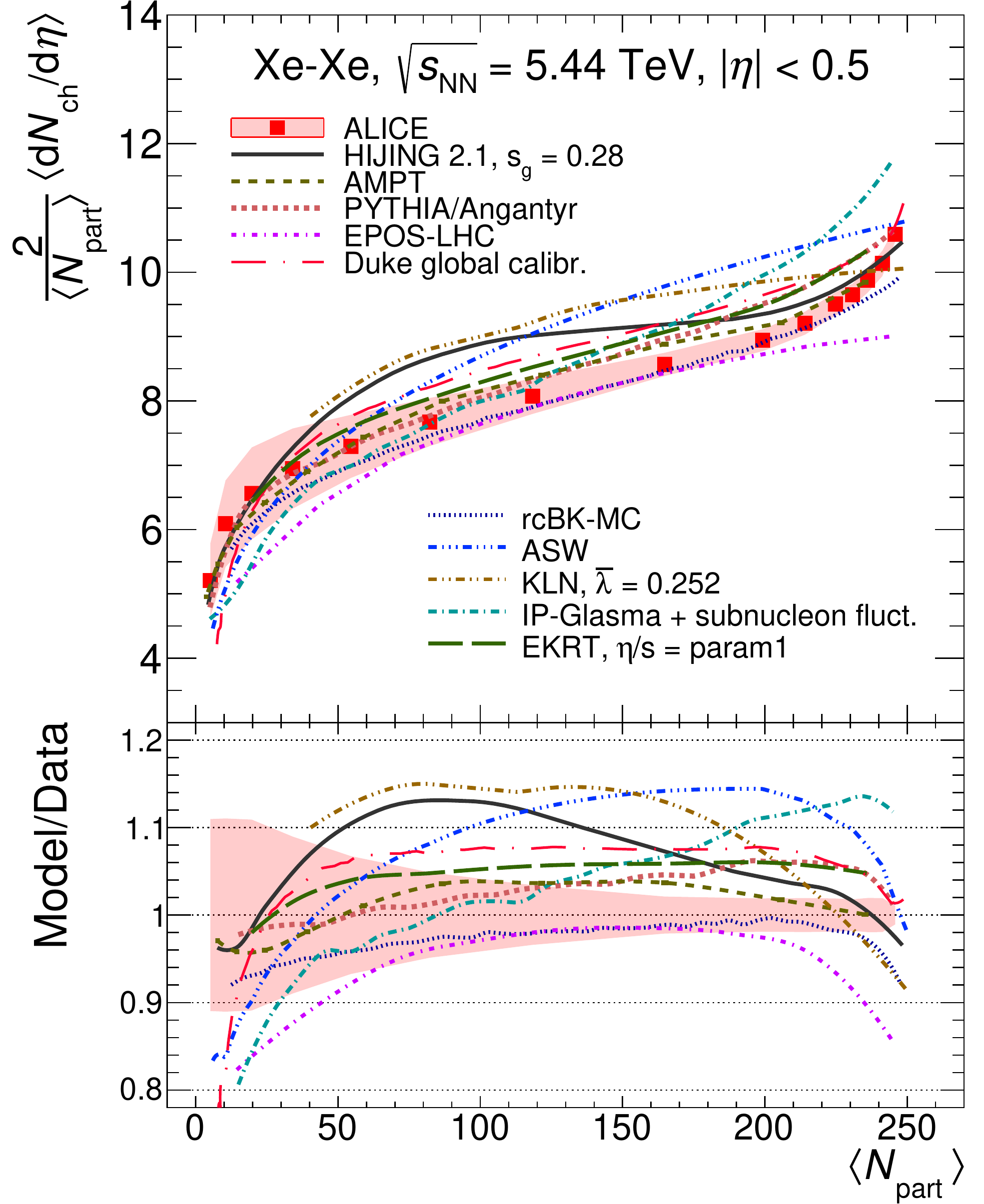}
      \caption{The \dNdetape\ vs \avNpart\ in Xe--Xe collisions~\cite{Acharya:2018hhy}}
      \label{fig:npartmodel}
    \end{subfigure}
  \end{center}
  \caption {\dNdetape\ in AA and pA collisions.}
\end{figure}

Figure \ref{fig:roots} shows $\dNdetape$ for the top 5\% central
Xe--Xe collisions compared to previous measurements for AA collisions as a
function of $\sqrt{s_\mathrm{NN}}$, as well as for inelastic and Non-Single Diffractive (NSD) pp,
$\mathrm{p\bar{p}}$ and NSD pA and dA collisions. The
lines are power law fits to the data.
For the 0--5\% central AA collisions, the Xe--Xe result fits with the power-law
trend previously established by the various measurements.
It indicates that Xe still acts like a heavy ion
compared to the pp and pA trends that are far from the central AA
collisions. The results in pA collisions for NSD are overlaid with the pp and
p$\mathrm{\bar{p}}$ trend of INEL collisions. This can be interpreted as if the
contribution from diffractive processes is negligible in pA collisions.
Figure \ref{fig:npart} shows \dNdetape\ in $|\eta|<0.5$ as a function of
\avNpart\ for various collision systems and collision energies. The distributions
of Xe--Xe at $\sqrt{s_\mathrm{NN}}=5.44$ TeV and Pb--Pb at
$\sqrt{s_\mathrm{NN}}=5.02$ TeV decrease by a factor of 2 from the most central to
peripheral collisions and smoothly connect to the results in pp and p--Pb
collisions. A steep rise of the distribution starting from $\avNpart=200$
for Xe-Xe collisions is newly observed and is thought to result from multiplicity fluctuations
due to the smaller size of Xe nucleus than heavier ones like Pb or Au.
This can be supported from the trend for Cu--Cu collisions in Fig \ref{fig:npart}.
The $\avNpart$-dependence of \dNdetape\ for Xe--Xe collisions is also compared with
theoretical models in Fig \ref{fig:npartmodel}. All the models describe the data
within $\pm 20\%$.

\section{Conclusions}

%All the results for charged-particle multiplicity density in LHC Run 1 and 2
%energies that ALICE has measured in inclusive and multiplicity dependent pp
%collisions, inclusive p--Pb collisions  and centrality dependent Pb--Pb and
%Xe--Xe collisions are provided in this document.
An overview of results for charged-particle multiplicity density in LHC Run 1 and 2 energies
measured by ALICE in p–-Pb, Pb–-Pb and Xe–-Xe collisions is provided in this proceedings.
% The MPI scenarios
%for the collective behaviour expect \dndeta\ distributions
%etter in high multiplicity pp collisions.
Newly measured \dndeta\ in
Xe--Xe collisions at $\sqrt{s_\mathrm{NN}}=5.44$ TeV shows still a heavy-ion like
behaviour.
All theoretical models based on various particle production mechanisms and different
initial conditions describe \dndeta\ vs $\eta$ and \dNdetape\ vs
\avNpart\ within $\pm20\%$ in pA and AA collisions.
This study might provide further constraints on models and help to improve
our understanding of the evolution of particle production with energy and system size.

%converged within $\pm20\%$ through LHC Run 1 and 2 periods.
%However,
%still more constraints are demanded  to converge the scenarios given in the
%theoretical models for the particle-production mechanism.

%% The Appendices part is started with the command \appendix;
%% appendix sections are then done as normal sections
%% \appendix

%% \section{}
%% \label{}

%% References
%%
%% Following citation commands can be used in the body text:
%% Usage of \cite is as follows:
%%   \cite{key}         ==>>  [#]
%%   \cite[chap. 2]{key} ==>> [#, chap. 2]
%%

%% References with BibTeX database:

\bibliographystyle{elsarticle-num}
\bibliography{nupha_bkkim}
%% Authors are advised to use a BibTeX database file for their reference list.
%% The provided style file elsarticle-num.bst formats references in the required Procedia style

%% For references without a BibTeX database:

% \begin{thebibliography}{00}

%% \bibitem must have the following form:
%%   \bibitem{key}...
%%

% \bibitem{}

% \end{thebibliography}

\end{document}